\documentclass[aps,pre,twocolumn,superscriptaddress,floats,floatfix]{revtex4-2}
\usepackage{amssymb,amsmath,amsthm}
\usepackage{graphicx}
\usepackage{epstopdf}
\usepackage{color}
\usepackage{subfigure}
\usepackage{epsfig}
\usepackage{rotating}
\usepackage{mwe}
\usepackage{float}
\usepackage{dcolumn,bm}
\usepackage{verbatim}
\usepackage{hyperref}
\usepackage{pgf,tikz}
\usepackage{comment}
\usepackage[normalem]{ulem}
\definecolor{Darkred}{RGB}{180, 0, 0}
\definecolor{Darkblue}{RGB}{50,116,181}
\hypersetup{
  colorlinks = true,     
  urlcolor   = Darkblue, 
  linkcolor  = Darkblue, 
  citecolor  = Darkred   
           }

\begin{document}

\begin{abstract}

We show that the Gross-Pitaevskii equation coupled with the wave equation for a wire (GP-W) provides a natural theoretical framework for understanding recent experiments employing a nanowire to detect a single quantum vortex in superfluid $^4 {\rm He}$. We uncover the complete spatiotemporal evolution of such wire-based vortex detection via direct numerical simulations of the GP-W system. Furthermore, by computing the spatiotemporal spectrum, we obtain the vortex-capture-induced change in the oscillation frequency of the wire. We quantify this frequency shift by plotting the wire's oscillation frequency versus time and obtain results that closely match experimental observations. In addition, we provide analytical support for our numerical results by deriving the dispersion relation for the oscillating wire, with and without a trapped vortex. We show that the Magnus force opens a gap in the wire dispersion relation. The size of the gap becomes the characteristic frequency of the wire when a vortex is trapped.
\end{abstract}

\title{Capture and release of quantum vortices using mechanical devices in low-temperature superfluids}
\author{Sanjay Shukla}
\email{shuklasanjay771@gmail.com}
\affiliation{Centre for Condensed Matter Theory, Department of Physics, Indian Institute of Science, Bangalore 560012, India}
\author{Giorgio Krstulovic}
\email{giorgio.krstulovic@oca.eu}
\affiliation{Universite C$\hat{o}$te d’Azur, Observatoire de la C$\hat{o}$te d’Azur, CNRS, Laboratoire Lagrange, Nice, France}
\author{Rahul Pandit $^1$}
\email{rahul@.iisc.ac.in}
\maketitle

Vortex filaments are rapidly rotating thin localized structures in fluid flows. They play an important role in our understanding of such flows and their turbulence~\cite{majda2002vorticity,chorin2013vorticity}. They appear as the dissipative structures of turbulent flow, and their dynamics has fascinated physicists and mathematicians for centuries~\cite{Thomson_1880}. Vortex filaments are also crucial in quantum fluids, such as superfluid helium and atomic BECs, as they are the most fundamental hydrodynamic excitation. Given their quantum origin, they are topological defects of the macroscopic wavefunction that describes the superfluid; the circulation of these defects assumes values that are multiples of the quantum of circulation {$\mathcal{K} =h/m$}, where $h$ is the Planck constant and $m$ the mass of the atoms. 

The characterization and detection of such vortices is a challenge for condensed-matter science. The main difficulty is the vanishingly small core size, which is of the order of one Angstrom in superfluid helium. 
Since the early experiments by Packard \cite{Packard_1969}, experimentalists have made enormous progress in the detection
of quantum vortices. Efforts to track and trap single quantum vortices have been made by groups that employ particles whose linear size is of the orders of micrometers, so it is considerably greater than the radius of the vortex core~\cite{nature_bewley,LaMantia_QuantumTurbulenceVisualized_2014,peretti2023direct}. Such experiments have studied vortex reconnections~\cite{Bewley_2008} and Kelvin waves~\cite{Fonda_DirectObservationKelvin_2014}, and they have carried out a direct and accurate verification of Feynman's rule that relates the vortex density to the rotation rate~\cite{peretti2023direct}. 

Recently, a different experimental technique has been pioneered by \emph{Guthrie, et al.}~\cite{natcomm_Guthriee}{, which has been successful in millikelvin superfluid $^4\rm He$ for studiying signatures of the interaction, entrapment, and release of quantum vortices from nanobeams.} These experiments rely on measurements of the frequency of the nanobeam and its modifications in time~\cite{natcomm_Guthriee}. These changes in time have been interpreted as the trapping and release of a quantum vortex by the nanobeam; however, so far no theoretical or numerical 
support has been presented in support of this interpretation.

\begin{figure*}[!hbt]
    \centering
    \includegraphics[scale=0.26]{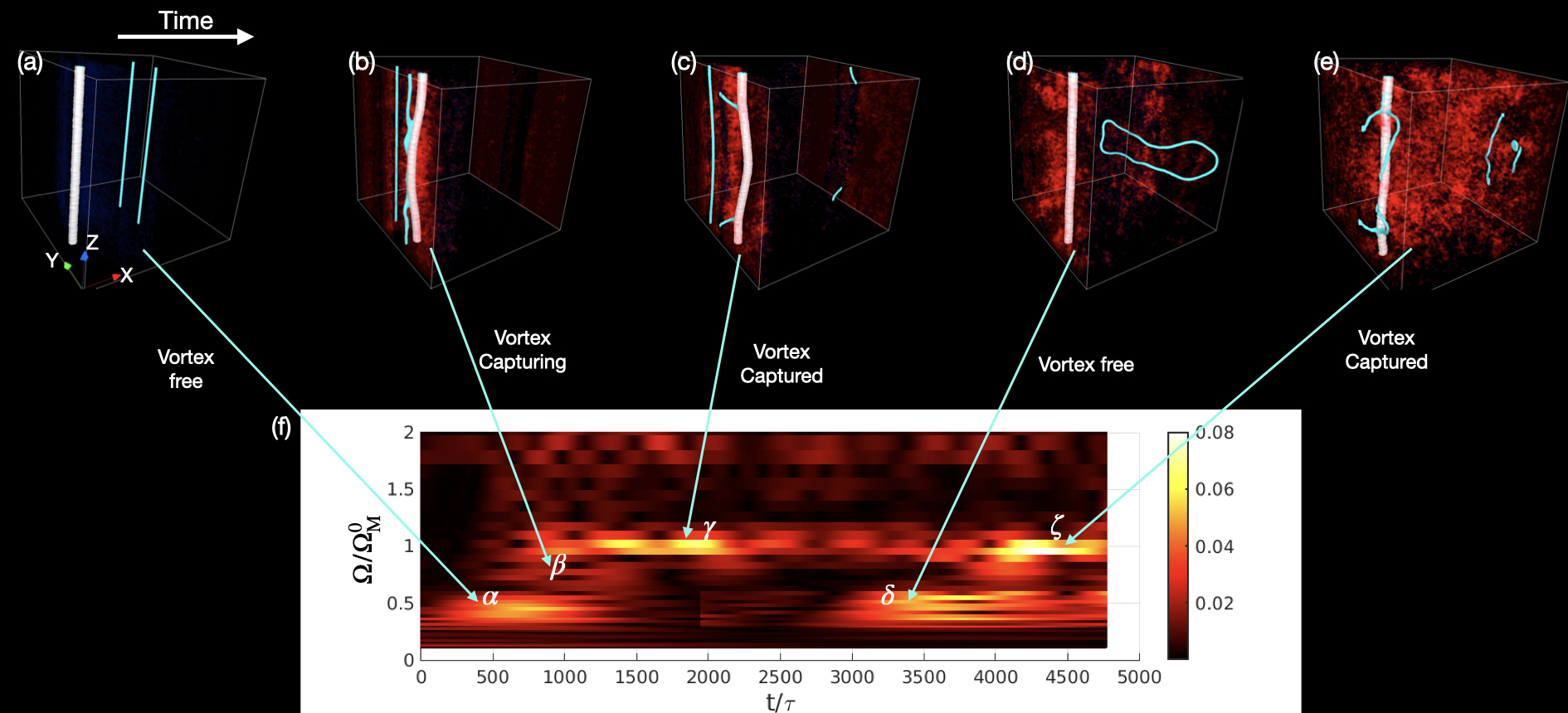}
    \caption{Isosurface plots of the density $|\psi|^2$ showing the spatiotemporal evolution of two quantum vortices (in cyan) interacting with the oscillating wire (in grey), which is depicted via the isosurface of the wire's potential $V_w$, in our GP-W model (see text), at five representative times that increase from left to right: (a) vortex-free state labelled by $\alpha$; (b) vortex-capturing state $\beta$; (c) vortex-captured state $\gamma$; (d) vortex-disentangled state $\delta$; and (e) vortex-recaptured state $\zeta$. These isosurface plots are superimposed on volume plots of $|\psi|^2$, in which sound waves appear as red clouds. (f) Filled contour plot of the absolute value of the continuous wavelet transform (cwt), which we obtain via the windowed Fourier transform in MATLAB, of the wire's position ${\bf q}=(q_x,q_y)$ as a function of the normalized frequency $\Omega/\Omega_{\rm M}^0$ of the wire and time $t/\tau$. The noteworthy changes in $\Omega/\Omega_{\rm M}^0$ are associated with the configurations in (a), (b), (c), (d), and (e) are labelled $\alpha$, $\beta$, $\gamma$, $\delta$, and $\zeta$, respectively.}
    \label{fig:vortex_wire_freq_time}
\end{figure*}

In this Letter, we design a model, based on the Gross-Pitaevskii equation (GPE), that is able to reproduce the results of the experiments of \emph{Guthrie, et al.}~\cite{natcomm_Guthriee}. We show explicitly how the natural frequencies of a wire, immersed in the superfluid, are modified by the trapping and release of a quantum vortex. Figure \ref{fig:vortex_wire_freq_time} summarizes our results via flow visualizations and the evolution of the temporal spectra. We also provide a theory that explains the changes in wire frequencies when a vortex is trapped on it. 

Our theoretical model uses the GPE to describe the superfluid via a complex macroscopic wavefunction $\psi$ and a solid deformable thin cylindrical object, immersed in this superfluid, by an external potential. Our approach is similar to that used in the modelling of large particles in superfluids~ \cite{Vishwanath_2018,Giuriato2018InteractionBA}. In particular, the cylinder is described by classical degrees of freedom and, for simplicity, we assume that the dynamics of the cylinder is given by the wave equation~\footnote{More sophisticated mechanical models, such as for a rectangular beam, can be obtained by a straightforward generalization of our model.}. We refer to our model as the Gross-Pitaevskii-Wire model (GP-W). The self-consistent coupled equations are derived by varying the total action of the system (see Sec.~\ref{sec:hamiltonian} in Appendix). The GP-W system of equations is:
\begin{eqnarray}
    i\hbar \frac{\partial \psi}{\partial t} &=& -\frac{\hbar^2}{2m} \nabla^2 \psi + g|\psi|^2 -\mu \psi + V_{\rm ext}(\bf r-R) \psi \,; \label{eq:GPE}\\
    \Ddot{\bf q} &=& -\frac{1}{\rho_w}\int dx dy  V_w \nabla \ |\psi(x,y,\zeta)|^2 \nonumber \\
    &+& c_w^2 \frac{\partial^2 \bf q}{\partial \zeta^2}\,;
    \label{eq:wave_equation}
\end{eqnarray}
here {${\bf r} = \{x,y,z\}$} is the 3D spatial coordinate, $g > 0$ is the strength of the interaction between the bosons, $\mu$ the chemical potential, and $V_w$ the potential that accounts for the wire, whose centerline is parametrized by ${\bf R}(\zeta)\equiv \{q_x(\zeta),q_y(\zeta),\zeta\}$, {where $\zeta$ is the parameter that varies along the central axis of the wire}. Equation~(\ref{eq:wave_equation}) governs the dynamics of the wire, with  ${\bf q}\equiv(q_x, q_y)$ the transverse displacement of the wire of density per unit length $\rho_w$, and speed of sound $c_w$. {Specifically, Eq.~(\ref{eq:wave_equation}) describes a vibrating string with the linear dispersion relation}. We model the wire by a Gaussian potential 
\begin{eqnarray}
    V_{\rm ext}({\bf r-R})= \lim_{\epsilon\to 0}\int V_w \frac{e^{-\frac{(z-\zeta)^2}{2\epsilon^2}}}{\sqrt{2\pi \epsilon^2}}d\zeta\,,
    \label{eq:wire_potential_external}
\end{eqnarray}
with
\begin{eqnarray}
    V_w = V_0 \exp\bigg(-\frac{( x - q_x(\zeta))^2+( y-q_y(\zeta))^2}{2a^2}\bigg)\,, 
    \label{eq:wire_potential_main}
\end{eqnarray}
where $V_0$ and $a$ are, respectively, its strength 
and width. { In Eq.~\eqref{eq:wire_potential_external}, the Gaussian function along $z$-direction, which becomes the delta function as its width $\epsilon\to 0$, is used to derive Eq.~\eqref{eq:wave_equation} from the Hamiltonian [see Sec.~\ref{sec:hamiltonian} in Appendix]}.  The GP-W system conserves the number of particles $N=\int |\psi|^2 d{\bf r}$, the total energy $E$, and the total momentum ${\bf P}$ [see Sec.~\ref{sec:hamiltonian} in Appendix].

The wire, with mass $M_w^0$, length $L_w$, and radius $a_w$, displaces some superfluid, whose (added) mass is $M_w = m \int (|\psi_{\infty}|^2 - |\psi_w|^2 )d{\bf r}$, where $\psi_{\infty}$ and $\psi_w$ are, respectively, the steady-state wavefunctions 
without and with the wire. {The radius of the wire is calculated using the displaced mass as $a_w = (\tfrac{3M_w}{\rho 4\pi})^{1/3}$, where $\rho \sim |\psi_{\infty}|^2$}. The mass ratio and effective mass 
\begin{eqnarray}
    \mathcal{M} \equiv \frac{M_w^0}{M_w}\;\; {\rm{and}} \qquad M_w^{\rm eff} = M_w^0+M_w
    \label{eq:added_mass}
\end{eqnarray}
help us to distinguish between light ($\mathcal{M}<1$) and heavy ($\mathcal{M}>1$) wires. The frequency of the wire, when a vortex is trapped on it, is obtained by a balance of its inertia and the Magnus force~\cite{Giuriato_2020} and is
\begin{eqnarray}
    \Omega_{\rm M}^0 = \frac{\rho \Gamma L_w}{M_{w}^{\rm eff}}\,,
    \label{eq:natural_freq}
\end{eqnarray}
where {$\Gamma=n_vh/m$} is the quantized circulation with $n_v$ being the multiplicity of the vortex and $\rho$ is the superfluid density.

Models of the type (\ref{eq:GPE}-\ref{eq:wave_equation}) account naturally for the processes of vortex trapping and release by solid boundaries, without recourse to any \textit{ad hoc} rule, so they have been used with great success, numerically and theoretically, to study the dynamics of particles in superfluids~\cite{Vishwanath_2018}, the interaction of particles and quantum vortices~\cite{Giuriato2018InteractionBA,Giuriato_2020,Griffin_MagnusforceModelActive_2020}, vortex reconnections with trapped particles~\cite{Giuriato_QuantumVortexReconnections_2020}, and active Lagrangian superfluid turbulence~\cite{Giuriato_ActiveFinitesizeParticles_2020}. The GPE is derived for weakly interacting Bose-Einstein condensates (BECs). Even though $^4 {\rm He}$ is a strongly interacting system, GPE-type models reproduce its observed hydrodynamical behavior well, at least qualitatively. Therefore, our GP-W model provides a powerful framework for the development of a theoretical understanding of the experiments~\cite{natcomm_Guthriee} that use nano-mechanical devices to study vortex dynamics in 
superfluid $^4 {\rm He}$.

To study Eqs.~(\ref{eq:GPE})-(\ref{eq:wire_potential_main}), we perform direct numerical simulations (DNSs)  via a Fourier pseudospectral method on a periodic cubical domain of side $L=2\pi$, with $N^3=256^3$ or $N^3=512^3$ collocation points, and the $2/3$-rule for dealiasing~\cite{giorgio_2011,HOU_dealiasing,Shukla_axion_2024}, i.e., spectral truncation of Fourier modes
for wavenumbers $k=|{\bf k}| > k_{max}=N/3$. In our DNSs, lengths are normalized by the healing length $\xi$ and time by $\tau=\xi/c_s$. We set the speed of sound to be $c_s=1$; and the healing length is $\xi = 2dx$, where $dx=L/N$. For time marching, we use the fourth-order Runge-Kutta scheme RK4 for Eq.~\eqref{eq:GPE} and the exponential time differencing RK4 scheme~\cite{cox2002exponential} for Eq.~\eqref{eq:wave_equation}, because the ratio $c_w/c_s$ is large. In our study, the radius and the length of the wire are, respectively, $a_w \simeq 3 {\xi} $ and  $L_w \simeq 256\xi$; the corresponding numbers in the nano-beam experiments~\cite{natcomm_Guthriee} with $^4{\rm He}$ are $a_w \simeq 100 {\xi}$ and $L_w \simeq 10^4{\xi}$. Thus, the ratio $a_w/L_w$ in our DNSs is comparable to that in experiments, so our study should be able to capture the essential properties of the wire-vortex interactions that have been observed in these experiments~\cite{natcomm_Guthriee}.

We use an initial condition $\psi_{vw}$ (see Sec.~\ref{sec:init_con} in Appendix), in which a vortex-anitvortex pair is positioned away from the wire; furthermore, we include a small perturbation on ${\bf q}$. As time $t$ progresses, the wire oscillates, the vortex-antivortex pair moves towards the wire, one of them gets captured by it, whereupon the frequency of the wire changes. An illustrative sequence of vortex capture, disentanglement, and recapture events is shown via pseudocolor plots in Figs.~\ref{fig:vortex_wire_freq_time}(a)-(e). In Fig.~\ref{fig:vortex_wire_freq_time}(f), we show 
filled contour plot of the absolute value of the continuous wavelet transform (cwt) of the wire's position ${\bf q}=(q_x,q_y)$ as a function of the normalized frequency $\Omega/\Omega_{\rm M}^0$ of the wire and time $t/\tau$. Note the changes in $\Omega/\Omega_{\rm M}^0$ that are associated with the configurations in (a), (b), (c), (d), and (e) that are labelled $\alpha$, $\beta$, $\gamma$, $\delta$, and $\zeta$, respectively. Clearly, our DNS reproduces the observed experimental behaviors of the nanobeam in Ref.~\cite{natcomm_Guthriee}; and it confirms that the changes in $\Omega/\Omega_{\rm M}^0$ arise from vortex trapping or disentangling. Furthermore, we observe the frequency of the wire, when the vortex is trapped on it, corresponds to the one given by the Magnus force~\eqref{eq:natural_freq}. It is also important to note that, when the vortex is not trapped, the frequency of the wire is $\simeq 0.4\Omega_{\rm M}^0$, which depends on the initial perturbation around the mean position of the wire and corresponds to the frequency of the dominant mode of vibration. 

We now develop a theory that allows us to understand the modification of the wire's natural frequency because of its interaction with the vortex.

In the absence of quantized vortices and with $V_w=0$, we can linearize the GPE in Eq.~(\ref{eq:GPE}) about the uniform equilibrium state $|\psi|^2=n_0$ to obtain the Bogoliubov dispersion~\cite{bogoliubov1947theory} relation 
\begin{eqnarray}
    \Omega_{\rm B}(k) = c_s |{\bf k}| \bigg(1+\frac{1}{2}\xi^2 |{\bf k}|^2\bigg)^{1/2}\,,
    \label{eq:Bogoliubov_dis}
\end{eqnarray}
where $c_s=\sqrt{gn_0/m}$ is the speed of sound, ${\bf k}$ is the wave vector of the excitation, and the healing length $\xi=\hbar/\sqrt{2gmn_0}$,  which yields the vortex-core radius. Fluctuations of quantum vortices include Kelvin waves (KW). The frequency spectrum $ \Omega_{\rm KW}$ for Kelvin waves was derived by Kelvin for a hollow vortex in an incompressible Euler fluid~\cite{Thomson_1880}:
\begin{eqnarray}
    \Omega_{\rm KW}(k) = \frac{\Gamma}{2\pi a_0^2}\Bigg[1-\sqrt{1+a_0k\frac{K_0(a_0k)}{K_1(a_0k)}}\Bigg]\,,
    \label{eq:Kelvin_wave_Om}
\end{eqnarray}
where $K_n(x)$ is the order-$n$ modified Bessel function and $k$ is the magnitude of the wave vector. Roberts~\cite{Roberts_2003} showed that the dispersion relation in Eq.~\eqref{eq:Kelvin_wave_Om} is also valid for a quantum vortex in GPE at small wavenumbers and obtained $a_0\simeq1.1265\xi$. Because of the lack of a full analytical dispersion relation of vortex excitations in the GPE, we use the following fit to the dispersion relation in our simulations~\cite{Giuriato_2020}:
\begin{eqnarray}
    \Omega_{\rm v}^{\rm fit} = \Omega_{\rm KW}[1+\epsilon_{\frac{1}{2}}(a_0k)^{\frac{1}{2}}+\epsilon_{1}(a_0k)+\frac{1}{2}(a_0k)^{\frac{3}{2}}]\,,
    \label{eq:omega_fit}
\end{eqnarray}
where $\epsilon_{\frac{1}{2}}=-0.20$ and $\epsilon_{1}=0.64$ are dimensionless parameters. For small wavenumber, $\Omega_{\rm v}^{\rm fit} \to \Omega_{\rm KW}$, while for large wavenumber, $\Omega_{\rm v}^{\rm fit} \propto k^2$, which is a free-particle dispersion relation, applicable for small-scale excitations of a superfluid vortex.

\begin{figure}[h!]
    \centering
    \includegraphics[scale=0.22]{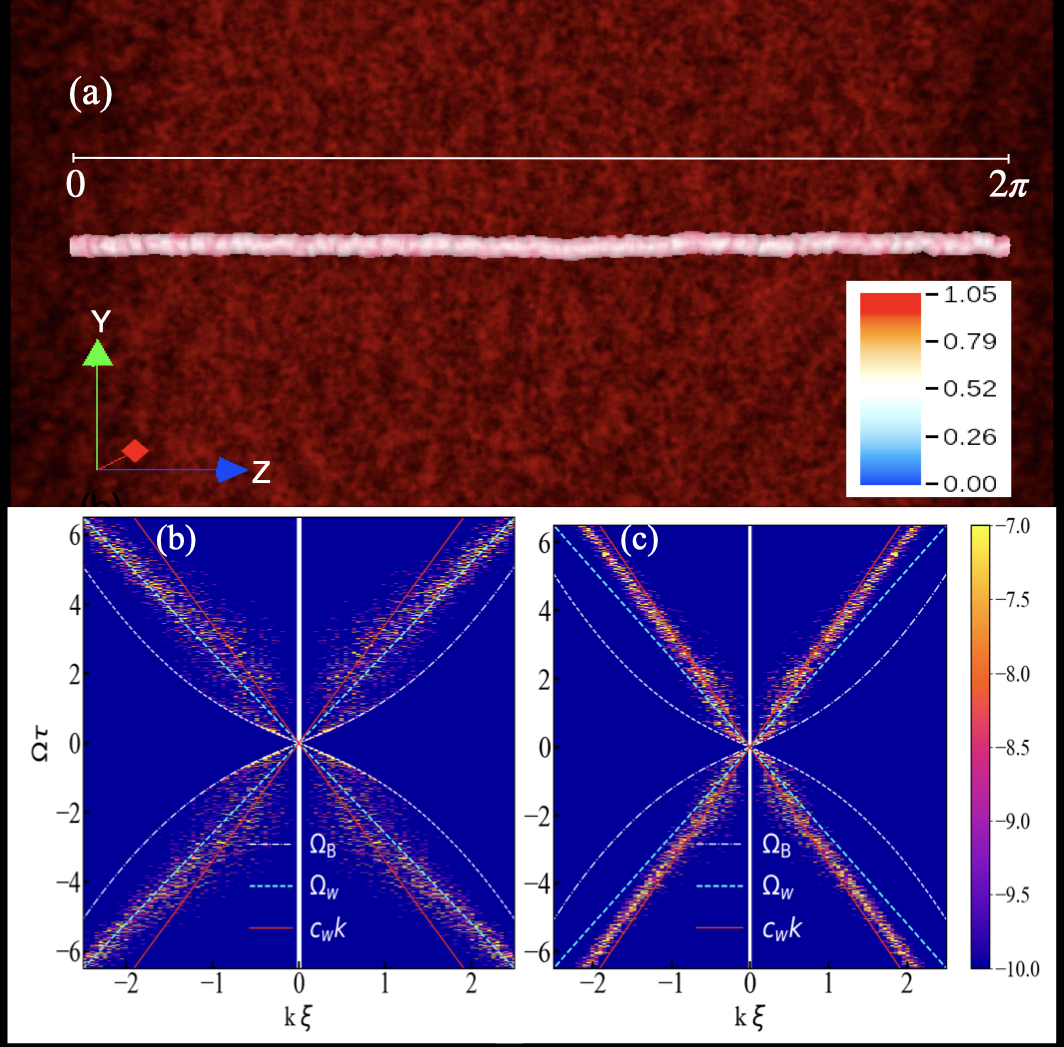}
    \put(-190,109){\color{white} $\mathcal{M}=2.0$}
    \put(-88,108){\color{white} $\mathcal{M}=5.0$}
    \caption{(a) One level contour plot of wire immersed in GP-W system, showing small-amplitude oscillations; this is
    superimposed on a volume plot of $|\psi|^2$. (b) Filled contour plot of the spatiotemporal spectrum $S_q(k,\omega)$ (see text) of the wire { on log scale};
    the cyan dashed line represents the linear dispersion relation of Eq.~(\ref{eq:omega_wire_lin}) with  $\mathcal{M}=2$, $c_s/c_w=1/3.3$, and the radius of the wire $a_w = 2.50\xi$. { The white dashed line shows the Bogoliubov dispersion curve of Eq.~\eqref{eq:Bogoliubov_dis}; and the sold red line shows the dispersion relation $c_wk$ without accounting for added mass effects}. (c) is the same as (b) but for $\mathcal{M}=5$ {with other parameters fixed}.}
    \label{fig:rho_wire_M6}
\end{figure}

If we consider the wire in the superfluid without vortices, then, in the long-wavelength limit, the dispersion relation for
the wire frequency is~\cite{Fetter_1971}
\begin{eqnarray}
    \Omega^{\pm}_w(k) \sim \pm c_w k \bigg(\frac{1}{1+\frac{1}{\mathcal{M}}}\bigg)^{1/2}\,, \qquad k \to 0,
    \label{eq:omega_wire_lin}
\end{eqnarray}
where $c_w$ is the speed of sound in the wire and the effect of the superfluid appears via $\mathcal{M}$, which accounts for the added mass [Eq.~\eqref{eq:added_mass}]. { The dispersion relation in Eq.~\eqref{eq:omega_wire_lin} is the solution of Eq.~\eqref{eq:wave_equation} in the long-wavelength limit. In the following paragraph, we verify it using our DNS of the GP-W model}. To study the wire's dispersion in Eq.~\eqref{eq:omega_wire_lin} without a vortex, we use an initial condition $\psi_w$ (see Sec.~\ref{sec:init_con_Fig2} in Appendix) in which all the modes of the wire are excited with a small amplitude { such that the amplitude is inversely proportional to the wavenumber $k$. So, the lowest mode is excited with the highest amplitude (see Sec.~\ref{sec:init_con_Fig2} in Appendix))}.  In our DNS for GP-W model, we obtain the spatiotemporal spectrum $S_q(k,\omega)\sim |q(k,\omega)|^2$, where $q$ is the magnitude of the displacement of the wire [see Eq.~\eqref{eq:wave_equation}]. In Fig.~\ref{fig:rho_wire_M6}(a) we show a representative plot of the wire (in gray) superimposed on a volume plot of $|\psi|^2$, whose fluctuations indicate sound waves. The dispersion relation~\eqref{eq:omega_wire_lin} is in good agreement with the low-$k$ part of high-intensity regions in the filled contour plot of $S_q(k,\omega)$ [Figs.~\ref{fig:rho_wire_M6}(b) and (c) for two values of $\mathcal{M}=2$ and $\mathcal{M}=5$]; these high-intensity regions yield the complete dispersion relation for all values of $k$.

\begin{figure}[h!]
    \centering
    \includegraphics[scale=0.21]{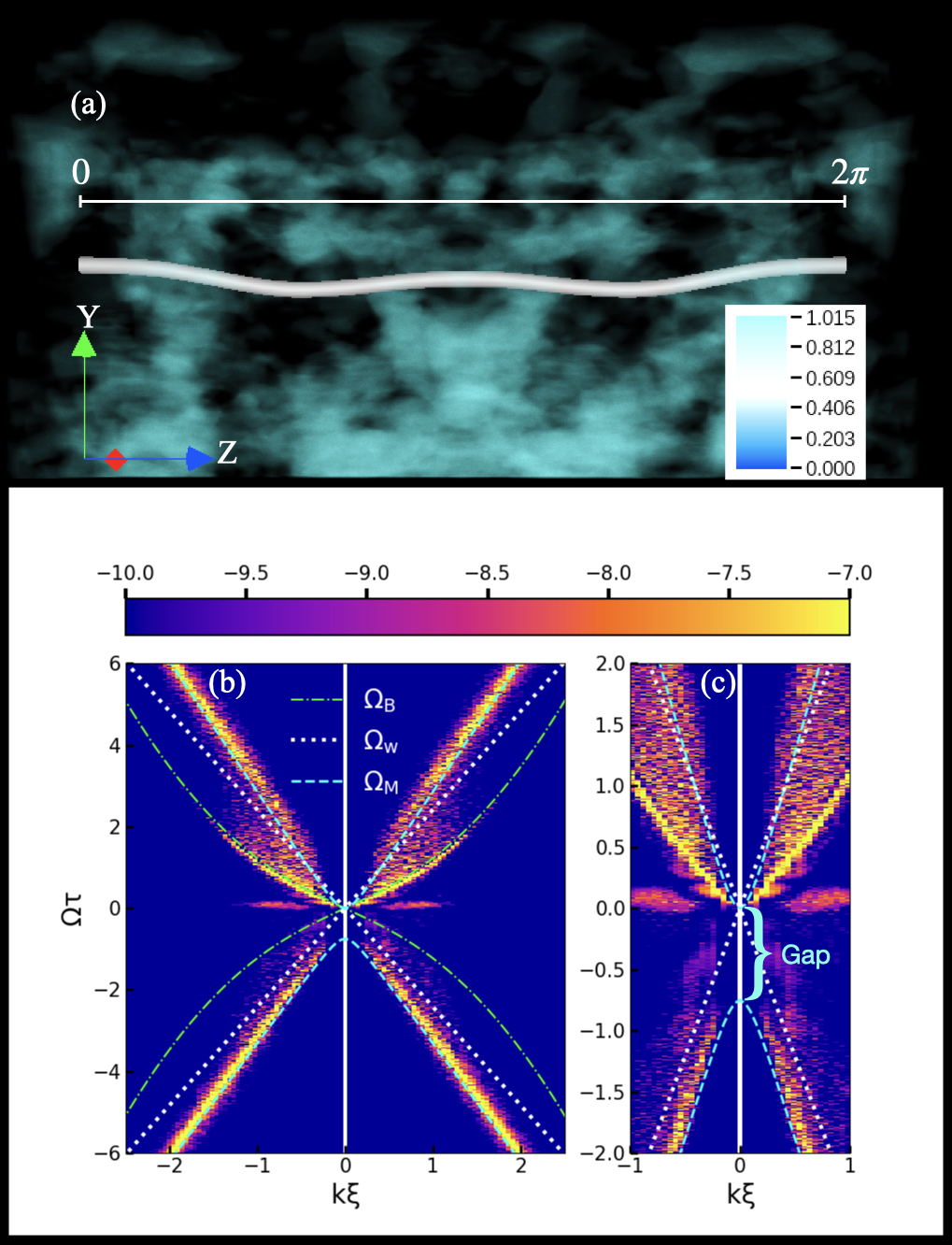}    
    \caption{One level contour plots depicting the wire in grey loaded with a vortex (a) at an intermediate time. The cyan cloud in (a)  represents the sound waves, which is a volume plot of $|\psi|^2$. (b) The spatiotemporal spectrum $S_q(k,\omega)\sim |q(k,\omega)|^2$ of the wire on a log scale placed at the location of a bare vortex, featuring small-amplitude Kelvin waves. The white dashed curve represents the dispersion relation from Eq.~(\ref{eq:Bogoliubov_dis}), whereas the cyan dashed line corresponds to the analytical dispersion relation in Eq.~(\ref{eq:omega_wire_vortex}). { The green dot-dashed line shows the Bogoliubov dispersion curve of Eq.~\eqref{eq:Bogoliubov_dis}}. (c) A closer examination of (b) that focuses on the central region. Here we have used $\mathcal{M}=4.0$, $c_s/c_w=1/2.7$, and the radius of the wire is $a_w=1.84\xi$.}
    \label{fig:vortex_wire_sp_M3.5}
    \end{figure}

To study how the wire's dispersion relation changes when a vortex is trapped on it, we use an initial condition with a perturbed wire enrobed by a vortex (see Sec.~\ref{sec:initial_condition_wire_vortex} in Appendix). For convenience, we chose $\Gamma<0$, so that Kelvin-wave excitations lie in the upper part of the dispersion relation. The measured dispersion relation for this system is presented in Fig.~\ref{fig:vortex_wire_sp_M3.5}(b).
Remarkably, the vortex modifies the excitationm spectrum of the wire; in particular, it lifts the degeneracy, at $k=0$, as is apparent in Fig.~\ref{fig:vortex_wire_sp_M3.5}(c). { We chose a different set of parameters in Fig.~\ref{fig:vortex_wire_sp_M3.5}(b) as compared to that in Fig.~\ref{fig:rho_wire_M6}(b) so that the gap in Fig.~\ref{fig:vortex_wire_sp_M3.5}(b) is much more enhanced leaving the physics unchanged. The measured dispersion relation matches with the analytical dispersion form in Eq.~\eqref{eq:omega_wire_vortex}, which we derive in the following section.}

To understand this perturbed spectrum, we derive an effective equation that governs the dynamics of the wire in the presence of a vortex. 
{First, we write the wire displacement in complex form $q=q_x+ iq_y$. Then, following the ideas of reference \cite{Giuriato_2020}, a simple equation for $q$ is obtained by balancing the Magnus force $i\Omega^0_{\rm M}(\dot{q}-v_{\rm si})$, the tension $c_w^2 \partial_{zz}q$, and wire inertia. Here, $v_{\rm si}$ is the self-induced velocity at position $q$, which depends on the model.} 
{For the sake of simplicity, we first employ the local-induction approximation (LIA)~\cite{Arms_1965}, in which the superfluid flow velocity can be approximated by $v_{\rm si} = i\frac{\Gamma}{4\pi} \Lambda \frac{\partial^2}{\partial z^2} {\bf R}(z,t)$, where
$\Lambda$ is a constant in this approximation of the order of $\log(\ell/\xi)$, with $\ell$ the inter-vortex distance. Note that in the LIA framework, Kelvin waves have the dispersion relation 
$\Omega_{\rm LIA} = -\tfrac{\Gamma \Lambda k^2}{4\pi}$.} The dynamics of the wire is governed by 
\begin{eqnarray}
    \Ddot{q}(z,t) &=& i\Omega_{\rm M}^0\bigg[\dot{q}(z,t)-i\frac{\Gamma}{4\pi} \Lambda \frac{\partial^2}{\partial z^2}q(z,t)\bigg]+c_w^2 \frac{\partial^2 }{\partial z^2}q(z,t)\,.\nonumber\\
    \label{eq:wire_equation_model}
\end{eqnarray}
{The previous equation is a natural generalization of the model introduced in~\cite{Giuriato_2020}}.

{We now seek solutions of Eq.~(\ref{eq:wire_equation_model}) of the form} $q(z,t) = q_0e^{i(\Omega_{\rm M}^{\pm}t-kz)}$ and then obtain their frequency
\begin{eqnarray}
    \Omega_{\rm M}^{\pm}(k) = \frac{\Omega_{\rm M}^0}{2}\pm \frac{1}{2} \sqrt{(\Omega_{\rm M}^0)^2+ \bigg(\frac{\Omega_{\rm M}^0\Gamma \Lambda}{\pi}+4c_w^2\bigg )k^2}\,.
    \label{eq:omega_wire_vortex_old}
\end{eqnarray}

We can go beyond the LIA by following Ref.~\cite{Giuriato_2020} to obtain the following 
dispersion relation for the waves along the wire:
\begin{eqnarray}
    \Omega_{\rm M}^{\pm}(k) = \frac{\Omega_{\rm M}^0}{2}\pm \frac{1}{2} \sqrt{(\Omega_{\rm M}^0)^2-4\Omega_{\rm M}^0\Omega_v^{\rm}+4c_w^2 k^2 }\,,
     \label{eq:omega_wire_vortex}
\end{eqnarray}
where $\Omega_v$ is a model-dependent bare vortex frequency. If we use the LIA, 
$\Omega_v= \Omega_{\rm LIA}$, so Eq.~\eqref{eq:omega_wire_vortex} becomes Eq.~\eqref{eq:omega_wire_vortex_old}; to go beyond the LIA we use $\Omega_v= \Omega_{\rm v}^{\rm fit}$, which is given by Eq.~\eqref{eq:omega_fit}. In the absence of a vortex, the wire's frequency spectrum (\ref{eq:omega_wire_lin}) exhibits a gapless mode with $\Omega^{\pm}_{ w} (k=0) = 0$. 
This mode develops a gap, i.e., $\Omega^+_{M} (k=0) - \Omega^-_{M} (k=0) = \Omega_{\rm M}^0 > 0$, when a vortex is introduced [see Eqs.~\eqref{eq:omega_wire_vortex} and Fig.~\ref{fig:vortex_wire_sp_M3.5}(c)]. The gap observed in our GP-W simulations is well explained by our theoretical prediction. Furthermore, the whole dispersion relation is reproduce by our theory \textit{without any adjustable parameter}, when we use the bare Kelvin-wave dispersion-relation fit $\Omega_{\rm v}^{\rm fit}$.

\noindent \textit{Conclusions:} Superfluid turbulence is often envisioned as a tangle of quantum vortices~\cite{Feynman_1995,schwarz1982generation,donnelly1988superfluid,pof_L_Skrbek,pnas_Carlo,pnas_L_Sk,paoletti2011quantum,barenghi2023quantum}. It is important, therefore, to detect such vortices unambiguously, which has been the goal of several experiments~\cite{Packard_1969,nature_bewley,Bewley_2008,Minowa_2022,Nago_PhysRevB_2013,natcomm_Guthriee,peretti2023direct}. The use of wires or nanobeams~\cite{Nago_PhysRevB_2013,natcomm_Guthriee} has led to very promising results that have shown how a vortex interacts with a wire. We have carried out a GPE study that has been designed to uncover such wire-vortex interactions and is, therefore, of direct relevance to the experiments of Refs.~\cite{Nago_PhysRevB_2013,natcomm_Guthriee}. In particular, we have obtained various signatures of the capture of a superfluid vortex by a wire using our DNS of the GP-W system. In the absence of any quantized vortices in the system, the wire oscillates with a frequency characterized by the initial random disturbance. We quantify how the wire-vortex interaction changes the oscillation frequency of the wire. In particular, we monitor the temporal evolution of the normalized frequency $\Omega/\Omega_{\rm M}^0$ to identify [Fig.~\ref{fig:vortex_wire_freq_time}] of vortex-free, vortex-trapping, and vortex-trapped states as in the experiments of Ref.~\cite{natcomm_Guthriee}. Our computation of 
the spatiotemporal spectrum allows us to obtain dispersion relations [Figs.~\ref{fig:rho_wire_M6}(b) and \ref{fig:vortex_wire_sp_M3.5}(b)] without and with vortices on the wire. {In particular, without a vortex, the transverse displacements of the wire, along two perpendicular directions, lead to a degeneracy in the excitation spectra [Figs.~\ref{fig:rho_wire_M6} and~\ref{fig:vortex_wire_sp_M3.5}]. If the vortex is present, a Magnus force acts on the wire because of the circulation of the vortex; this lifts the degeneracy in the frequency spectrum and leads to a gap. This Magnus force produces two circularly polarised modes that are separated by the frequency $\Omega_{\rm M}^0$.}.
This frequency thus becomes dominant for sufficiently large-scale perturbations, explaining the jumps observed on the inset of Fig.\ref{fig:rho_wire_M6}.
We show that our DNS results are in good agreement with our simple linear theory. Finally, our model is easily generalizable to describe different types of mechanical devices, or the dynamics of several devices connected by vortices. 

\begin{acknowledgments}
We thank the Indo-French Centre for Applied Mathematics (IFCAM), the Science and Engineering Research Board (SERB), and the National Supercomputing Mission (NSM), India for support, and the Supercomputer Education and Research Centre (IISc) for computational resources. GK acknowledges the financial support of Agence Nationale de la Recherche through the project QuantumWIV ANR-23-CE30-0024-02 and the Simons Foundation Collaboration grant Wave Turbulence (Award No. 651471).
SS thanks GK for his hospitality during his research visit to the Observatoire de la C\^ote d'Azur, Nice, France.
\end{acknowledgments}

\appendix
\section{The GP-W Hamiltonian}
\label{sec:hamiltonian}
The GP-W system, comprising a weakly interacting superfluid and a vibrating wire, is governed by the following Hamiltonian:
\begin{equation}
\begin{aligned}
    \mathcal{H} &= \int_{\mathcal{V}} d{\bf r} \left[ \frac{\hbar^2}{2m} 
  |\nabla \psi({\bf r},t)|^2 + \frac{1}{2}g |\psi({\bf r},t)|^4 - \mu|\psi({\bf r},t)|^2 \right. \\
  & \left. +  V_{\rm ext}({\bf r-R}) |\psi({\bf r},t)|^2 \bigg]  + \int dz \bigg[ \frac{{\bf p}^2}{2\rho_w} + \frac{1}{2} c_w^2 \rho_w \bigg(\frac{\partial {\bf q}}{\partial z}\bigg)^2\right]\,.
  \label{eq:Hamiltonian}
  \end{aligned}
\end{equation}
The external potential, $V_{\rm ext}$, of the wire is given by the following
\begin{eqnarray}
    V_{\rm ext}({\bf r-R})= \int V_w \frac{e^{-\frac{(z-\zeta)^2}{2\epsilon^2}}}{\sqrt{2\pi \epsilon^2}}d\zeta\,,
    \label{eq:wire_potential_ext}
\end{eqnarray}
with
\begin{eqnarray}
    V_w = V_0 \exp\bigg(-\frac{( x - q_x(\zeta))^2+( y-q_y(\zeta))^2}{2a^2}\bigg)\,, 
    \label{eq:wire_potential}
\end{eqnarray}
where $V_0$ and $a$ are, respectively, its strength 
and width. The above Hamiltonian conserves the number of particles $N=\int |\psi|^2 d{\bf r}$ and the total energy $E = E_{\rm GP} + E_{\rm w}$, where $E_{\rm GP}$ and $E_{\rm w}$ are the total energies contained in the GPE and wire respectively

\begin{eqnarray}
E_{\rm GP} &=& \int d{\bf r}\bigg[\frac{\hbar^2}{2m} |\nabla \psi|^2 +\frac{1}{2}g |\psi|^4 - \mu |\psi|^2 \nonumber \\
&+& V_w({\bf r-R}) |\psi|^2 \bigg] \,, \nonumber \\
E_{\rm w} &=& \int dz \bigg[ \frac{1}{2} \rho_w |\dot{\bf q}|^2 +\frac{1}{2} c_w^2 \rho_w \left| \frac{\partial {\bf q}}{\partial z}\right|^2 \bigg]
\label{eq:total_energy_gp}
\end{eqnarray}

The Hamiltonian~(\ref{eq:Hamiltonian}) also conserves the total momentum ${\bf P} = {\bf p}_{\rm GP} + {\bf p}$, where ${\bf p}_{\rm GP}$ and  ${\bf p}$ are the momenta contained in GPE and the wire, respectively:
\begin{eqnarray}
    {\bf p}_{\rm GP} &=& \frac{i\hbar}{2} \int d{\bf r} \bigg(  \psi \nabla \psi^*-\psi^* \nabla \psi \bigg)\,, \nonumber \\
    {\bf p} &=& \rho_w \int dz \  \dot{\bf q}(z)\,.
    \label{eq:total_momentum}
\end{eqnarray}

\section{Initial condition for Fig.1: wire away from the vortex}
\label{sec:init_con}
To get a stationary state, $\psi_w$, of the Gross-Pitaevskii (GP) system with the wire in it (without any vortex), we start with the initial mean position of the wire ${\bf q}_i = (q_{xm},q_{ym})$ and solve the following imaginary-time version of Eq.~\eqref{eq:GPE} in the main text:
\begin{equation}
\begin{aligned}
    \hbar \frac{\partial \psi}{\partial t} = \frac{\hbar^2}{2m} \nabla^2 \psi - g|\psi|^2 +\mu \psi - V_w(\bf r-R) \psi \,.
    \label{eq:Imag_GPE}
    \end{aligned}
\end{equation}
We now use the Pad\'e approximation~\cite{Berloff_2004} for the vortex solution of the GP system. 
If $\rho(r)$ is the density distribution of a vortex in the Pad\'e approximation, the vortex-antivortex pair wavefunction is 
\begin{eqnarray}
    \psi_{vp} = \sqrt{\rho(|{\bf r}-{\bf r}_1|)}\cdot \sqrt{\rho(|{\bf r}-{\bf r}_2|)}\cdot e^{i\theta_1}\cdot e^{i\theta_2}\,,
    \label{eq:Pade}
\end{eqnarray}
where ${\bf r}_1$ and ${\bf r}_2$ are the positions of the vortices in the $xy$ plane, and $\theta_1$ and $\theta_2$ are their corresponding phases. The initial distance between the two vortices is $\pi/2$. We use $\psi_{vp}$ as the initial condition and solve Eq.~(\ref{eq:Imag_GPE}) with $V_w=0$ to minimize the sound waves. Finally the initial condition for the GP-W is $\psi_{vw} = \psi_w\cdot \psi_{vp}$.

For Fig.~\ref{fig:vortex_wire_freq_time} (f), we construct a complex vector $q=(q_x,q_y)$ from the wire's displacements. This vector is given to the MATLAB function ``cwt" which stands for continuous wavelet transformation. The function ``cwt" uses the vector $q$, the window function $W$, and the sampling frequency $F_s$ as the input. The ``cwt" performs the Fourier transform over a window $W$ and gives the spectrum with frequency as a function of time.

\section{Initial condition for Fig.2: wire without the vortex}
\label{sec:init_con_Fig2}
We first get the stationary state, $\psi_w$, of the Gross-Pitaevskii system with wire in it, using Eq.~\eqref{eq:Imag_GPE}. The wire is now perturbed such that all the modes have an equal amount of energy, which we can using the total energy of the wire as
\begin{eqnarray}
    E_w =   \int dz \bigg[ \frac{{\bf p}^2}{2\rho_w} + \frac{1}{2} c_w^2 \rho_w \bigg(\frac{\partial {\bf q}}{\partial z}\bigg)^2\bigg]\,.
\end{eqnarray}
We can write the above equation in Fourier space using Parseval's theorem as follows
\begin{eqnarray}
    E_w^k =   \int dk \bigg[ \frac{|{\bf p}^k|^2}{2\rho_w} + \frac{1}{2} c_w^2 \rho_w k^2 |{\bf q}^k|^2\bigg]\,,
\end{eqnarray}
where ${\bf q}^k$, ${\bf p}^k$ are the position and momentum in Fourier space, respectively, and $k$ is the wavenumber. From the equipartition theorem, each quadratic mode has an energy $\tfrac{k_{\rm B}T}{2}$, whence we get
\begin{eqnarray}
    |{\bf q}^k| = \sqrt{\frac{k_{\rm B}T}{\rho_w}} \frac{1}{c_wk}\nonumber\\
    |{\bf p}^k| = \sqrt{k_{\rm B}T\rho_w}\,,
\end{eqnarray}
where, $k_{\rm B}$ is the Boltzmann constant, $T$ is the temperature, $\rho_w$ is the line density of the wire, and $c_w$ is the speed of sound on the wire.

\section{Kelvin waves}
\label{sec:KW}
To show the effect of Kelvin waves, we start with four straight vortices of alternating signs to maintain the periodicity. We then solve the imaginary time ($t\to -i t$) version of the GPE [Eq.(1) in the main text], with $V_w = 0$ to obtain an equilibrium state by minimizing the emission of phonons (sound waves). We introduce a KW by perturbing a vortex as follows:
\begin{eqnarray}
    x(z) &=& A\sum_i^n \cos(iz+\phi_i^x)\,, \nonumber\\
    y(z) &=& A\sum_i^n \cos(iz+\phi_i^y)\,,
    \label{eq:KW}
\end{eqnarray}
where $\phi_i^x$, $\phi_i^y$ are random phases, $n=4$ is the number of Kelvin modes, and amplitude $A=1.5\xi$.

\section{Initial condition for Fig.3: wire at the position of a vortex}
\label{sec:initial_condition_wire_vortex}
We prepare an initial condition featuring four straight vortices using Eq.~\eqref{eq:Pade}, strategically positioned to maintain periodicity. The wire is situated at the location of one of these vortices. To introduce Kelvin waves, we initiate a disturbance in the vortex using Eq.~(\ref{eq:KW}) in Sec.~\ref{sec:KW}. In Fig.3(a) in the main text, the wire (illustrated in grey) is loaded with a vortex with Kelvin waves. We suppress the images of the other three vortices for clarity. As time progresses, the vortex becomes ensnared by the wire, leading to the initiation of wire oscillations coupled with the generation of sound waves, depicted as the cyan cloud in Fig.3(a). The frequency spectrum of the wire changes because of the trapped vortex in it; we show this by plotting the spatiotemporal spectrum $S_q(k,\omega)\sim |q(k,\omega)|^2$ of the wire in Fig.3(b). Notably, the degeneracy in the wire spectrum is decisively lifted at the origin (Fig.3(c)), and the spectrum is in agreement with the analytical frequency spectrum derived in Eq.(13).


\begin{thebibliography}{10}

\bibitem{majda2002vorticity}
A.~J. Majda, A.~L. Bertozzi, and A.~Ogawa, ``Vorticity and incompressible flow. cambridge texts in applied mathematics,'' {\em \href{ https://doi.org/10.1017/CBO9780511613203}{Appl. Mech. Rev.}}, vol.~55, no.~4, pp.~B77--B78, 2002.

\bibitem{chorin2013vorticity}
A.~J. Chorin, {\em Vorticity and turbulence}, vol.~103.
\newblock \href{https://doi.org/10.1007/978-1-4419-8728-0}{Springer Science \& Business Media}, 2013.

\bibitem{Thomson_1880}
W.~Thomson, ``Xxiv. vibrations of a columnar vortex,'' {\em \href{https://doi.org/10.1080/14786448008626912}{The London, Edinburgh, and Dublin Philosophical Magazine and Journal of Science}}, vol.~10, no.~61, pp.~155--168, 1880.

\bibitem{Packard_1969}
R.~E. Packard and T.~M. Sanders, ``Detection of single quantized vortex lines in rotating he ii,'' {\em \href{https://link.aps.org/doi/10.1103/PhysRevLett.22.823}{Phys. Rev. Lett.}}, vol.~22, pp.~823--826, Apr 1969.

\bibitem{nature_bewley}
G.~Bewley, D.~Lathrop, and K.~Sreenivasan, ``Visualization of quantized vortices,'' {\em \href{https://doi.org/10.1038/441588a}{Nature}}, vol.~441, 2006.

\bibitem{LaMantia_QuantumTurbulenceVisualized_2014}
M.~La~Mantia and L.~Skrbek, ``Quantum turbulence visualized by particle dynamics,'' {\em Physical Review B}, vol.~90, p.~014519, July 2014.

\bibitem{peretti2023direct}
C.~Peretti, J.~Vessaire, {\'E}.~Durozoy, and M.~Gibert, ``Direct visualization of the quantum vortex lattice structure, oscillations, and destabilization in rotating 4he,'' {\em \href{https://www.science.org/doi/abs/10.1126/sciadv.adh2899}{Science Advances}}, vol.~9, no.~30, p.~eadh2899, 2023.

\bibitem{Bewley_2008}
G.~P. Bewley, M.~S. Paoletti, K.~R. Sreenivasan, and D.~P. Lathrop, ``Characterization of reconnecting vortices in superfluid helium,'' {\em \href{https://www.pnas.org/doi/abs/10.1073/pnas.0806002105}{Proceedings of the National Academy of Sciences}}, vol.~105, no.~37, pp.~13707--13710, 2008.

\bibitem{Fonda_DirectObservationKelvin_2014}
E.~Fonda, D.~P. Meichle, N.~T. Ouellette, S.~Hormoz, and D.~P. Lathrop, ``Direct observation of {Kelvin} waves excited by quantized vortex reconnection,'' {\em Proceedings of the National Academy of Sciences}, vol.~111, pp.~4707--4710, Mar. 2014.

\bibitem{natcomm_Guthriee}
A.~Guthrie and et~al., ``Nanoscale real-time detection of quantum vortices at millikelvin temperatures,'' {\em \href{https://doi.org/10.1038/s41467-021-22909-3}{Nature Communications}}, vol.~12, 2021.

\bibitem{Vishwanath_2018}
V.~Shukla, R.~Pandit, and M.~Brachet, ``Particles and fields in superfluids: Insights from the two-dimensional gross-pitaevskii equation,'' {\em \href{https://link.aps.org/doi/10.1103/PhysRevA.97.013627}{Phys. Rev. A}}, vol.~97, p.~013627, Jan 2018.

\bibitem{Giuriato2018InteractionBA}
U.~Giuriato and G.~Krstulovic, ``Interaction between active particles and quantum vortices leading to kelvin wave generation,'' {\em \href{https://doi.org/10.1038/s41598-019-39877-w}{Scientific Reports}}, vol.~9, 2018.

\bibitem{Note1}
More sophisticated mechanical models, such as for a rectangular beam, can be obtained by a straightforward generalization of our model.


\bibitem{Giuriato_2020}
U.~Giuriato, G.~Krstulovic, and S.~Nazarenko, ``How trapped particles interact with and sample superfluid vortex excitations,'' {\em \href{https://link.aps.org/doi/10.1103/PhysRevResearch.2.023149}{Phys. Rev. Res.}}, vol.~2, p.~023149, May 2020.

\bibitem{Griffin_MagnusforceModelActive_2020}
A.~Griffin, V.~Shukla, M.-E. Brachet, and S.~Nazarenko, ``Magnus-force model for active particles trapped on superfluid vortices,'' {\em Physical Review A}, vol.~101, p.~053601, May 2020.

\bibitem{Giuriato_QuantumVortexReconnections_2020}
U.~Giuriato and G.~Krstulovic, ``Quantum vortex reconnections mediated by trapped particles,'' {\em Physical Review B}, vol.~102, p.~094508, Sept. 2020.

\bibitem{Giuriato_ActiveFinitesizeParticles_2020}
U.~Giuriato and G.~Krstulovic, ``Active and finite-size particles in decaying quantum turbulence at low temperature,'' {\em Physical Review Fluids}, vol.~5, p.~054608, May 2020.

\bibitem{giorgio_2011}
G.~Krstulovic and M.~Brachet, ``Energy cascade with small-scale thermalization, counterflow metastability, and anomalous velocity of vortex rings in fourier-truncated gross-pitaevskii equation,'' {\em \href{https://link.aps.org/doi/10.1103/PhysRevE.83.066311}{Phys. Rev. E}}, vol.~83, p.~066311, Jun 2011.

\bibitem{HOU_dealiasing}
T.~Y. Hou and R.~Li, ``Computing nearly singular solutions using pseudo-spectral methods,'' {\em \href{https://www.sciencedirect.com/science/article/pii/S0021999107001623}{Journal of Computational Physics}}, vol.~226, no.~1, pp.~379--397, 2007.

\bibitem{Shukla_axion_2024}
S.~Shukla, A.~K. Verma, M.~E. Brachet, and R.~Pandit, ``Gravity- and temperature-driven phase transitions in a model for collapsed axionic condensates,'' {\em \href{https://link.aps.org/doi/10.1103/PhysRevD.109.063009}{Phys. Rev. D}}, vol.~109, p.~063009, Mar 2024.

\bibitem{cox2002exponential}
S.~M. Cox and P.~C. Matthews, ``Exponential time differencing for stiff systems,'' {\em Journal of Computational Physics}, vol.~176, no.~2, pp.~430--455, 2002.

\bibitem{Berloff_2004}
N.~G. Berloff, ``Padé approximations of solitary wave solutions of the gross–pitaevskii equation,'' {\em \href{https://dx.doi.org/10.1088/0305-4470/37/5/011}{Journal of Physics A: Mathematical and General}}, vol.~37, p.~1617, jan 2004.

\bibitem{bogoliubov1947theory}
N.~Bogoliubov, ``On the theory of superfluidity,'' {\em J. Phys}, vol.~11, no.~1, p.~23, 1947.

\bibitem{Roberts_2003}
P.~H. Roberts, ``On vortex waves in compressible fluids. ii. the condensate vortex,'' {\em \href{https://doi.org/10.1098/rspa.2002.1033}{Proceedings of the Royal Society of London}}, vol.~459, p.~597–607, 2003.

\bibitem{Fetter_1971}
A.~L. Fetter and K.~Harvey, ``Elastic filaments and vortex waves,'' {\em \href{https://link.aps.org/doi/10.1103/PhysRevA.4.2305}{Phys. Rev. A}}, vol.~4, pp.~2305--2308, Dec 1971.

\bibitem{Arms_1965}
R.~J. Arms and F.~R. Hama, ``{Localized‐Induction Concept on a Curved Vortex and Motion of an Elliptic Vortex Ring},'' {\em \href{https://doi.org/10.1063/1.1761268}{The Physics of Fluids}}, vol.~8, pp.~553--559, 04 1965.

\bibitem{Feynman_1995}
R.~Feynman, ``Application of quantum mechanics to liquid helium. progress in low temperature physics,'' {\em \href{https://doi.org/10.1016/S0079-6417(08)60077-3}{Progr.LowTemp.Phys}}, vol.~1, pp.~17--53, 1955.

\bibitem{schwarz1982generation}
K.~Schwarz, ``Generation of superfluid turbulence deduced from simple dynamical rules,'' {\em \href{https://link.aps.org/doi/10.1103/PhysRevLett.49.283}{Physical Review Letters}}, vol.~49, no.~4, p.~283, 1982.

\bibitem{donnelly1988superfluid}
R.~J. Donnelly, ``Superfluid turbulence,'' {\em \href{https://www.jstor.org/stable/24989266}{Scientific American}}, vol.~259, no.~5, pp.~100--109, 1988.

\bibitem{pof_L_Skrbek}
L.~Skrbek and K.~R. Sreenivasan, ``{Developed quantum turbulence and its decay)},'' {\em \href{https://doi.org/10.1063/1.3678335}{Physics of Fluids}}, vol.~24, p.~011301, 01 2012.

\bibitem{pnas_Carlo}
C.~F. Barenghi, L.~Skrbek, and K.~R. Sreenivasan, ``Introduction to quantum turbulence,'' {\em \href{https://www.pnas.org/doi/abs/10.1073/pnas.1400033111}{Proceedings of the National Academy of Sciences}}, vol.~111, no.~supplement\_1, pp.~4647--4652, 2014.

\bibitem{pnas_L_Sk}
L.~Skrbek, D.~Schmoranzer, Šimon Midlik, and K.~R. Sreenivasan, ``Phenomenology of quantum turbulence in superfluid helium,'' {\em \href{https://www.pnas.org/doi/abs/10.1073/pnas.2018406118}{Proceedings of the National Academy of Sciences}}, vol.~118, no.~16, p.~e2018406118, 2021.

\bibitem{paoletti2011quantum}
M.~S. Paoletti and D.~P. Lathrop, ``Quantum turbulence,'' {\em \href{https://doi.org/10.1146/annurev-conmatphys-062910-140533}{Annu. Rev. Condens. Matter Phys.}}, vol.~2, no.~1, pp.~213--234, 2011.

\bibitem{barenghi2023quantum}
C.~F. Barenghi, L.~Skrbek, and K.~R. Sreenivasan, {\em Quantum turbulence}.
\newblock \href{ https://doi.org/10.1017/9781009345651}{Cambridge University Press}, 2023.

\bibitem{Minowa_2022}
Y.~Minowa, S.~Aoyagi, S.~Inui, T.~Nakagawa, G.~Asaka, M.~Tsubota, and M.~Ashida, ``Visualization of quantized vortex reconnection enabled by laser ablation,'' {\em \href{https://www.science.org/doi/abs/10.1126/sciadv.abn1143}{Science Advances}}, vol.~8, no.~18, p.~eabn1143, 2022.

\bibitem{Nago_PhysRevB_2013}
Y.~Nago, A.~Nishijima, H.~Kubo, T.~Ogawa, K.~Obara, H.~Yano, O.~Ishikawa, and T.~Hata, ``Vortex emission from quantum turbulence in superfluid ${}^{4}$he,'' {\em \href{https://link.aps.org/doi/10.1103/PhysRevB.87.024511}{Phys. Rev. B}}, vol.~87, p.~024511, Jan 2013.

\end{thebibliography}
\end{document}